\documentclass{appolb}
\usepackage{epsfig}
\begin{document}
% \eqsec  % uncomment this line to get equations numbered by (sec.num)
\title{Jets in ep and $\gamma$p Scattering at HERA%
\thanks{Presented at Photon2005, August 31 - September 4 2005, Warsaw, Poland,
on behalf of the H1 and ZEUS collaborations.}%
% you can use '\\' to break lines
}
\author{Thomas Sch\"orner-Sadenius
\address{Hamburg University, IExpPh, Luruper Chausse 149, 22761 Hamburg, Germany}
%\and
%the Name(s) of other Author(s)
%\address{and their affiliation}
}
\maketitle
\begin{abstract}
Recent jet physics results from ep and $\gamma$p scattering at HERA will be
reviewed covering cross-section measurements, the extraction of QCD
parameters, the transition region from photoproduction to deep-inelastic
scattering and the question of parton evolution in the proton.  
\end{abstract}
  
\section{Introduction}
Over the past few years, great progress has been achieved in the understanding
of the hadronic final state and of jet production at
HERA. This is, for example, expressed in the extremely good description of jet
cross-sections at high values of the photon virtuality, $Q^2$, or the precise
determinations of the strong coupling parameter, $\alpha_S$, from jet
data. 

At present, our best knowledge is contained in next-to-leading (NLO) order QCD
calculations which rely on collinear
factorisation and the use of DGLAP evolution. However, for some observables
the NLO precision does not seem to be sufficient - with many measurements now
being dominated by theoretical uncertainties higher orders in the
perturbative expansion are necessary. 
In addition, there are areas in jet physics at HERA which have not yet
been succesfully addressed with standard approach of DGLAP-based
factorisation, for example the question of the parton evolution scheme in
forward jet production. 

In this contribution, both the successes and the problems of jet physics at
HERA will be reviewed, based on the most recent results. 
%In section~\ref{cross-sections} jet cross-sections measurements in
%photoproduction and deep-inelastic scattering (DIS) and their comparisons to
%NLO theory will be shown. 
%In section~\ref{alphas}, extractions of the strong coupling
%parameter, $\alpha_S$, will be introduced together with the results of
%global QCD fits of the parton distribution functions
%(PDFs) using jet data in addition to the usual inclusive $F_2$ data. 
%Section~\ref{transition} highlights the problems of the
%transition region from photoproduction to DIS, and section~\ref{evolution}
%focuses on the question of parton evolution in the
%proton. 

%\begin{figure}[t]
%\begin{center}
%\epsfig{file=375_5c-1.eps,width=6.cm}
%\label{zeus-eps375-5}
%\caption{ZEUS inclusive jet cross-sections as function of transverse jet
%  energy in six regions of $Q^2$. Presented is the ratio (data-NLO)/NLO.}
%\end{center}
%\end{figure}

\section{Cross-section measurements in photoproduction and DIS}
\label{cross-sections}
Both the H1 and ZEUS collaborations have used almost all of the HERA-I data
sample of more than 80~${\rm pb^{-1}}$ for the measurement of inclusive, two-
and three-jet cross-sections in the high-$Q^2$ regime of $Q^2 >$~125~${\rm
  GeV^2}$ or similar of DIS~\cite{herajets}. 
The data are corrected to hadron level using
leading-order (LO) Monte Carlo (MC) generators and are compared to NLO QCD
calculations which have been corrected to the hadron level. The description of
the data by the theory is in general excellent; typically the data are
described within $\pm$5~$\%$, providing a handle on $\alpha_S$ and
the PDFs. 

%As an example,
%figure~\ref{zeus-eps375-5} shows the ratio (data-NLO)/NLO for the inclusive
%jet cross-section as function of the transverse jet energy in the Breit
%reference frame in six regions of $Q^2$ as measured by the ZEUS
%collaboration. Even for the bottom two plots, which corresond to the highest
%$Q^2$ values above 2000~${\rm GeV^2}$, most of the data points are described
%by the theory within 5~$\%$. 

Also in photoproduction the description of dijet
cross-sections which have newly been measured by the H1
collaboration~\cite{h1-photo} by the NLO theory including resolved photon
contributions is very good, making these data a candidate for further
constraining the photon PDFs. 

\begin{figure}[t]
\begin{center}
\epsfig{file=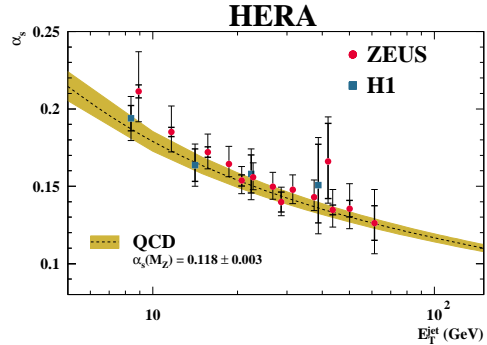,width=7.cm}
\caption{Summary of HERA measurements of $\alpha_S$ in comparison to the world
  average (dark band). The running of the strong coupling is clearly
  demonstrated by the HERA data.}
\label{claudia-plot1}
\end{center}
\end{figure}

\section{Extraction of $\alpha_S$ and the proton PDFs}
\label{alphas}
The high quality of the theoretical description of the DIS and photoproduction
jet cross-sections make these data ideal candidates for the extraction of the
strong coupling parameter, $\alpha_S$. For this enterprise, the dependence of
the NLO cross-section on $\alpha_S(M_Z)$ is parametrized by a quadratic
function, $\sigma^{theo}_i=A_i\cdot\alpha_S(M_Z)+B_i\cdot\alpha_S^2(M_Z)$, 
with the parameters $A_i$ and $B_i$ 
determined in a fit procedure individually for each observable $i$. 
Then, for each
observable $i$, the measured cross-section is mapped onto $\sigma^{theo}_i$,
leading 
to a value of $\alpha_S(M_Z)$ for each observable $i$. 
The various single values
of $\alpha_S(M_Z)$ can then be combined or evolved to the relevant
renormalisation scale in order to demonstrate the running of the QCD coupling. 
Figure~\ref{claudia-plot1}~\cite{claudia-hepex} summarizes the various
determinations of $\alpha_S$ from jet measurements from both H1 and ZEUS. The
data are consistent and agree very well with the world
average~\cite{bethke} indicated by the brown band, demonstrating the
running behaviour of the coupling. The most precise $\alpha_S$
measurements from HERA jets now have errors which are comparable to LEP
results. 
The various HERA measurements have been
used in~\cite{claudia-hepex} to derive a HERA average which is given as
$\alpha_S(M_Z) = 0.1186\pm 0.0011({\rm exp.})\pm 0.0050({\rm theo.})$.
The value is consistent with
the world average, the error is dominated by theory.

%\begin{figure}[t]
%\begin{center}
%\epsfig{file=gluon.eps,width=7.cm}
%\label{gluon}
%\caption{Fractional error on the gluon density as function of $x$ in bins of
%  $Q^2$~\cite{zeusfit}.}
%\end{center}
%\end{figure}

Recently, jet measurements~\cite{jetsforfit} in both photoproduction 
and DIS have been used by
the ZEUS collaboration in their global QCD fits in addition to the inclusive
structure function measurements with the specific aim of
further constraining the gluon density of the proton at high fractional
momenta $x$~\cite{zeusfit}. This technically challenging endeavour leads to a
massive improvement on the gluon PDF around $x =$~0.1.
%, see figure~\ref{gluon}
%which shows the fractional error on the gluon density as a function of $x$ in
%various regions of $Q²$ before (yellow) and after (red) the inclusion of jet
%data in the fit. 
A further benefit of the jet data is the improvement
of the $\alpha_S$ determination from the global fit. 

\section{The transition from photoproduction to DIS}
\label{transition}
\begin{figure}[t]
\begin{center}
\epsfig{file=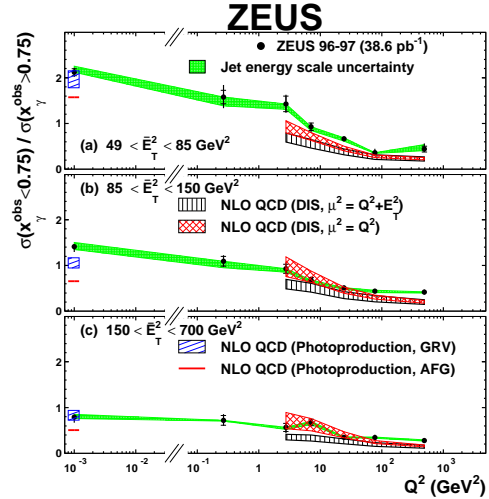,width=7.cm}
\caption{The ratio of the resolved over the direct dijet cross-section as
  function of $Q^2$ in three regions of the mean dijet $E_T$.}
\label{ratior}
\end{center}
\end{figure}

If the squared transverse energy of a parton, $E_T^2$, is smaller than the photon
virtuality $Q^2$, the parton is able to probe the hadronic substructure of
the photon. This effect is maximal for photoproduction and is more and more
suppressed for more and more virtual photons in DIS. Figure~\ref{ratior} from a
recent ZEUS measurement~\cite{desy-04-053} shows the ratio $R$ of the resolved
over the direct contribution to the dijet cross-section over a wide range of
photon virtualities from photoproduction to high-$Q^2$. It can be seen that
even at high values of $Q^2$ of 100~${\rm GeV^2}$ or higher, the resolved
component contributes significantly to the cross-section - an effect that is
not accounted for by the direct-photon only NLO QCD calculations which are
typically employed for DIS analyses. The concept of resolved contributions, on
the other hand, works very successfully for the photoproduction regime where
the data are well described by the NLO calculation which implements both
direct and resolved photon contributions.    

In a previous analysis~\cite{triple_analysis} H1 tried to better understand the
implications of the resolved photon structure
and to evaluate the necessary amount of resolved contributions to the cross-section
by studying the triple-differential cross-section
$d^3\sigma/dE_TdQ^2dx_{\gamma}$. Again, especially for $Q^2 <$~10~${\rm
  GeV^2}$ the data are undershot by the NLO prediction, even if the calculation
incorporates the resolved photon component. In contrast, leading-order
MC programs which incorporate parton shower algorithms and both longitudinally
and transversely polarised resolved photons describe the data rather well,
except for the highest $x_{\gamma}$ values. So, in conclusion, there are clear
indications for effects beyond direct NLO QCD - parton shower algorithms and
resolved photon contributions are necessary to describe the data. 

The H1 data have also been compared to the predictions of the CASCADE
program which incorporates the CCFM evolution and thus uses an
unintegrated gluon density. Especially in the medium $Q^2$ region between 10
and 25~${\rm GeV^2}$ the data are described by CASCADE; this might be
due to the fact that the unintegrated PDFs violate the strict $k_T$ ordering
of direct-photon DGLAP-type simulated events in much the same way that the
ordering is violated in resolved events. 
 
\section{The question of parton evolution scheme}
\label{evolution} 

\begin{figure}[t]
\begin{center}
\begin{minipage}{3.7cm}
\epsfig{file=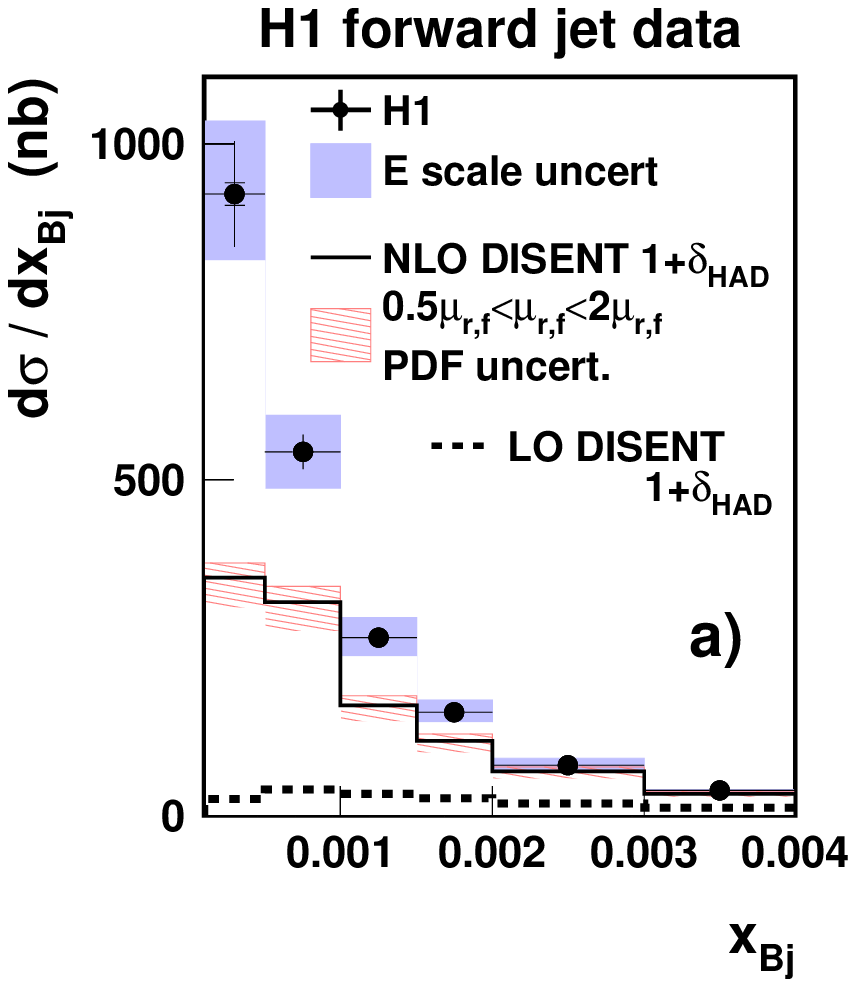,width=3.5cm}
\end{minipage}
\begin{minipage}{3.7cm}
\epsfig{file=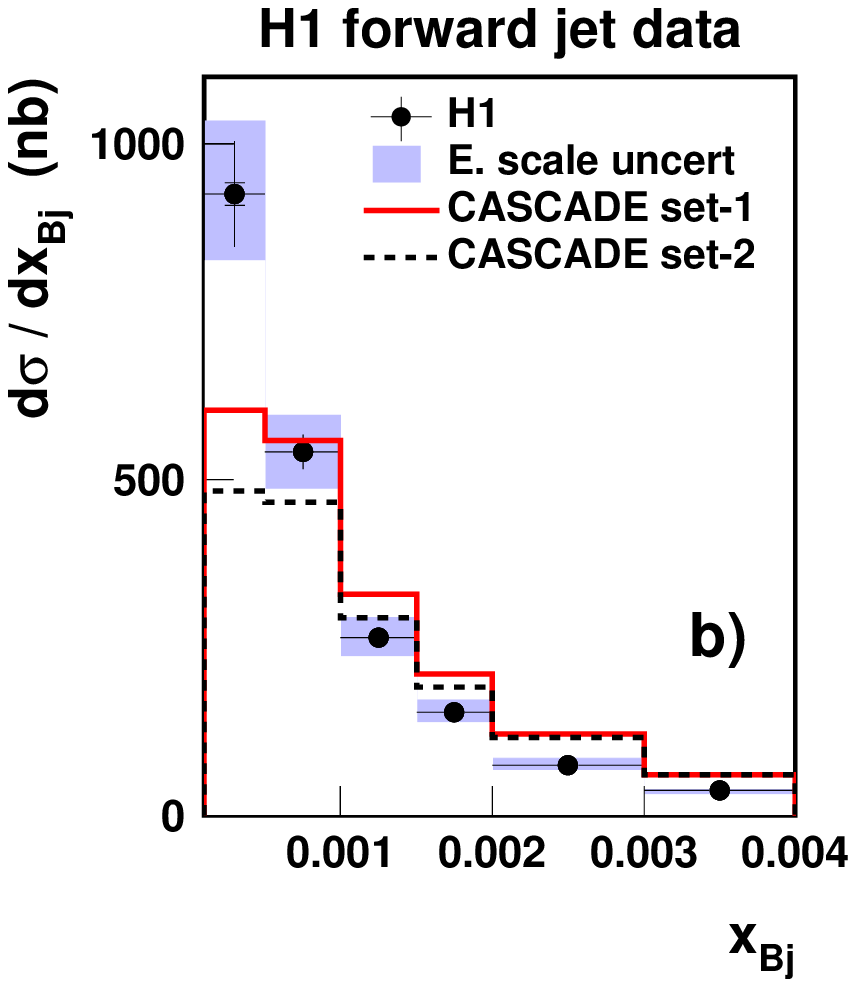,width=3.5cm}
\end{minipage}
\begin{minipage}{3.7cm}
\epsfig{file=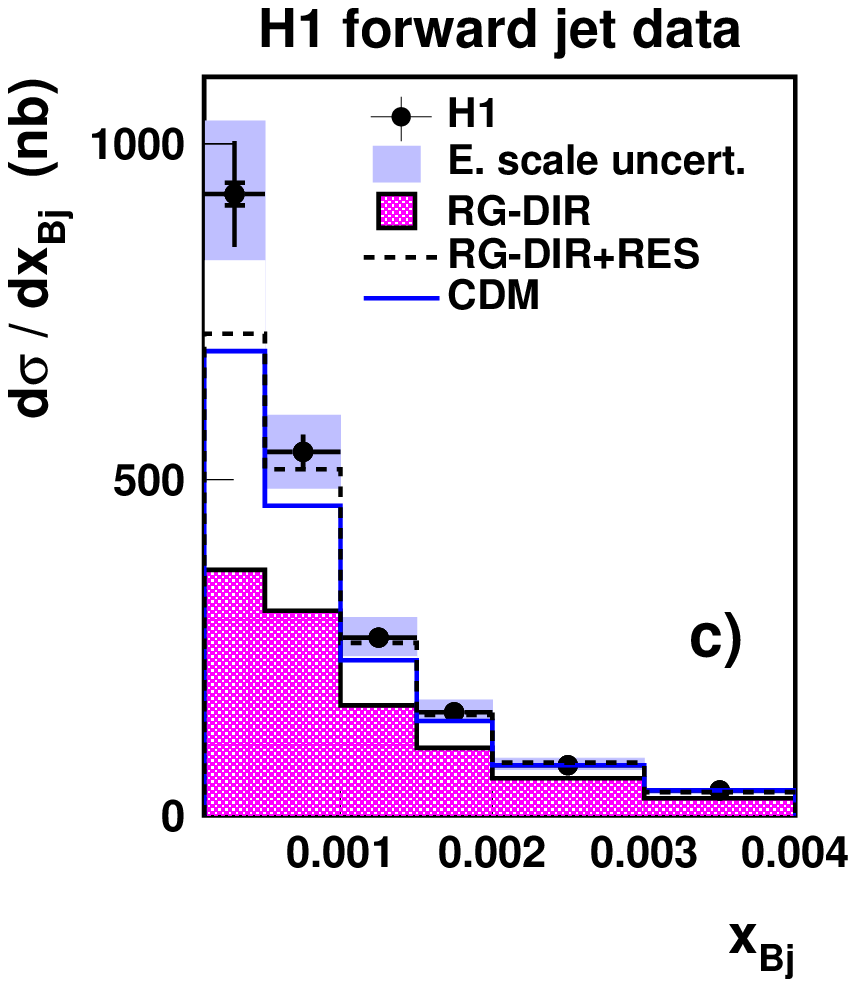,width=3.5cm}
\end{minipage}
\caption{H1 forward jet data~\cite{forwardjetsh1}, 
compared to fixed-order NLO calculations (left),
to the CCFM model CASCADE (center), and to DGLAP-type LO MC models (right).}
\label{h1fwdjets}
\end{center}
\end{figure}

The problem of parton evolution in the proton is addressed
by new analyses of forward jet events from both H1~\cite{forwardjetsh1} and 
ZEUS~\cite{forwardjetszeus}. 
Figure~\ref{h1fwdjets} shows the H1 result which
presents the data as function of $x_{Bj}$ and compares them to various models
and calculations. For both H1 and the very similar ZEUS result 
the fixed-order NLO calculation fails to describe the data at low $x_{Bj}$
values; in the case of ZEUS however, who in contrast to H1
use $Q^2$ as renormalisation
scale, the theoretical uncertainties on the predictions are huge. 
MEPS-type LO MC models also fail, although the inclusion of resolved
contributions brings them rather close to the data (see line `RG+DIR+RES' in 
figure~\ref{h1fwdjets}). For the ZEUS result the best description of the data   
is achieved by the ARIADNE MC which implements the color dipole model and thus
has similarity to the BFKL evolution scheme.

H1 has also measured the triple-differential cross-section in $x_{Bj}$, $Q^2$
and the transverse forward jet momentum $p_T$. 
Using the ratio of $p_T^2$ and $Q^2$
one can roughly adjust the type of evolution that is kinematically possible,
between DGLAP-resolved dominated ($Q^2 < p_T^2$), DGLAP-direct ($Q^2 > p_T^2$)
and BFKL-like ($Q^2 \sim p_T^2$). It turns out that both the CDM and the DGLAP
models work rather well in the kinematical areas for which they are considered 
applicable; programs using the CCFM evolution have problems in correctly
describing the shape of the distributions. 

The H1 measurement is also the first to show, for a selection of forward jets
plus a hard dijet system, a
discrepancy in the quality of the data description between the CDM model and a
direct+resolved DGLAP-based model, the latter failing to describe the data
where the CDM does a good job. This is interpreted as an amount 
of $k_T$ unordering in the resolved models that is not enough to account
for the forward jet features, in contrast to the CDM. 

All effects together clearly demonstrate 
that effects beyond the usual DGLAP-type direct-photon NLO picture are
present; this has also been pointed out by a further ZEUS
result~\cite{fwdjets2}. The fact that both BFKL-like models (CDM) and models
with resolved photon contributions give the best description of the data is
indicative for a violation of $k_T$ ordering in the process. However, the
success of the resolved models might also be attributed to their partial
simulation of higher orders in the perturbative expansion, mimicking parts of
a more complete (not yet available) NNLO calculation. 

The CASCADE program, which implements the CCFM
evolution, does not give a clear message from all above mentioned
measurements; in addition it depends strongly on the unintegrated gluon
density. There seem to be, however, indications in the H1
result~\cite{forwardjetsh1}  that CCFM is able to describe the data at least in a
specially designed (BFKL-like) phase-space. 

\section{Conclusion and outlook}

The overview given so far is clearly not complete, missing, for
example, measurements of inter-jet energy flow in photoproduction, the
analysis of color dynamics in DIS and photoproduction or the measurement of
subjet distributions from ZEUS~\cite{assortedstuff}. However, it should have
demonstrated that jet physics at HERA is at the same time a valuable
laboratory for precision QCD tests and an exciting place to study open
questions concerning, for example, the r${\hat{{\rm o}}}$le of resolved photon
contributions or the problem of the parton evolution in the proton. 

So far, the analysis of HERA-II data has hardly begun, most analyses being
restricted to about 80~${\rm pb^{-1}}$ or less from the HERA-I period. 
With already close to 150~${\rm pb^{-1}}$ from HERA-II and almost two more years of
HERA running ahead, one can clearly expect improvements in statistical precision.
But there is clearly hope to also improve
both the systematic and the theoretical uncertainties. Since the latter are by
now dominating many measurements, theoretical progress is of special
importance in this field. 

Finally it should be mentioned that the knowlegde gained in jet analyses at
HERA is valuable for the physics program of the LHC, as was demonstrated in
the HERA-LHC workshop. 

\section*{Acknowledgement}
I thank the organisers of Photon2005 for a stimulating
conference and my HERA colleagues for their support in preparing this
contribution.

\end{document}